\documentclass[aps,twocolumn,prl,reprint,amsmath,amssymb,superscriptaddress]{revtex4-1}

\usepackage{graphicx}
\usepackage{dcolumn}
\usepackage{multirow}
\usepackage{hhline}
\usepackage{booktabs}
\usepackage{bm}
\usepackage{microtype}
\usepackage{color}
\usepackage{xcolor}
\usepackage{epstopdf}

\begin{document}

\title{Valley subband splitting in bilayer graphene quantum point contact}

\author{R. Kraft}
\affiliation{Institute of Nanotechnology, Karlsruhe Institute of Technology, D-76021 Karlsruhe, Germany}

\author{I.V. Krainov}
\affiliation{A.F. Ioffe Physico-Technical Institute, 194021 St. Petersburg, Russia}

\author{V. Gall}
\affiliation{Institute of Nanotechnology, Karlsruhe Institute of Technology, D-76021 Karlsruhe, Germany}
\affiliation{Institute for Condensed Matter Theory, Karlsruhe Institute of Technology, D-76128 Karlsruhe, Germany}

\author{A.P. Dmitriev}
\affiliation{A.F. Ioffe Physico-Technical Institute, 194021 St. Petersburg, Russia}

\author{R. Krupke}
\affiliation{Institute of Nanotechnology, Karlsruhe Institute of Technology, D-76021 Karlsruhe, Germany}
\affiliation{Department of Materials and Earth Sciences, Technical University Darmstadt, Darmstadt, Germany}

\author{I.V. Gornyi}
\affiliation{Institute of Nanotechnology, Karlsruhe Institute of Technology, D-76021 Karlsruhe, Germany}
\affiliation{A.F. Ioffe Physico-Technical Institute, 194021 St. Petersburg, Russia}
\affiliation{Institute for Condensed Matter Theory, Karlsruhe Institute of Technology, D-76128 Karlsruhe, Germany}

\author{R. Danneau}\email[Author to whom correspondence should be addressed. Electronic mail: romain.danneau@kit.edu]{}
\affiliation{Institute of Nanotechnology, Karlsruhe Institute of Technology, D-76021 Karlsruhe, Germany}

\begin{abstract}

We report a study of one-dimensional subband splitting in a bilayer graphene quantum point contact in which quantized conductance in steps of $4\,e^2/h$ is clearly defined down to the lowest subband. While our source-drain bias spectroscopy measurements reveal an unconventional confinement, we observe a full lifting of the valley degeneracy at high magnetic fields perpendicular to the bilayer graphene plane for the first two lowest subbands where confinement and Coulomb interactions are the strongest and a 
peculiar merging/mixing of $K$ and $K'$ valleys from two non-adjacent subbands with indices $(N,N+2)$ which are well described by our semi-phenomenological model.

\end{abstract}

\maketitle

Thirty years after its discovery, quantized conductance resulting from the discretization of the one-dimensional (1D) subbands in a ballistic constriction remains one of the most striking effect in mesoscopic physics~\cite{Vanwees1988,Wharam1988,Beenakker1991,VanHouten1992,dattabook}. Thanks to the rapid development of nanofabrication, the quantum point contact (QPC) geometry \cite{Thornton1986} used in these experiments has become a basic tool to study 1D physics \cite{Giamarchibook} and design complex devices and circuits, as it can act as a beam splitter in electron-optics like experiments  \cite{Vanwees1989,Yacoby1994,Schuster1997,Henny1999,Comforti2002,Ji2003,Neder2007,Roulleau2007,Bieri2009,Bocquillon2013,Dubois2013,Jullien2014,Helzel2015,Banerjee2017} as well as noninvasive charge detectors \cite{Field1993,Sprinzak2002,Vandersypen2004,Gustavsson2006,Reilly2007,Cassidy2007} when the conductance is set below the first conductance plateau. While a vast majority of 1D ballistic systems shows quantized conductance in units of $2\,e^2/h$, where the factor of two is due to spin degeneracy, only few involve an additional valley degree of freedom such as Si-SiGe heterostructures \cite{Tobben1995,Wieser2002,Scappucci2006,Goswami2007,McGuire2010}, AlAs quantum wells \cite{Gunawan2006}, carbon nanotubes \cite{Biercuk2005} or single layer and bilayer graphene (SLG and BLG) \cite{Lin2008,Lian2010,Tombros2011,Allen2012,Goossens2012,Droscher2012,Terres2016,Li2016,Somanchi2017,Overweg2018,Caridad2018,Banszerus2018}.
Spin and valley degeneracy should give rise to a conductance of $4\,e^2/h$ per channel. However, deviations from the expected quantized conductance value have been mostly observed \cite{Tobben1995,Wieser2002,Scappucci2006,Goswami2007,McGuire2010,Gunawan2006,Biercuk2005,Lin2008,Lian2010,Tombros2011,Allen2012,Goossens2012,Overweg2018,Caridad2018,Banszerus2018}, and usually explained by the lifting of the valley degeneracy due to confinement.

	\begin{figure}
		\centering
		\includegraphics{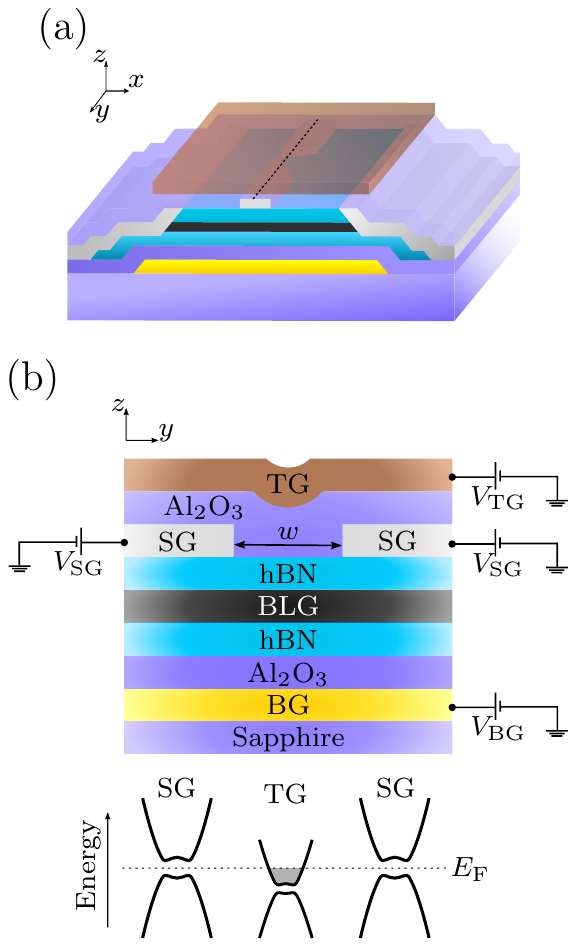}
		\caption{(a) Schematic illustrating the device layout. (b) Cross section of the device along the dashed line in (a), together with a spatial sketch of the electronic band structure across the constriction defined by the split gate.}
		\label{fig:1}
	\end{figure}

Controlling the valley isospin and breaking the valley degeneracy appears to be crucial in the development of valleytronics \cite{Schaibley2016}. Valley degeneracy could be tuned under various conditions and geometries \cite{Recher2007,Recher2009,Zhang2011,Knothe2018}; in graphene, the design of valley filters and valley valves have been proposed based on ballistic point contact \cite{Rycerz2007}. In addition, lifting the valley degeneracy appears to be essential in graphene spin qubit \cite{Trauzettel2007}. Here we present experiments on ballistic transport through a QPC electrostatically defined in BLG. To study the non-trivial splitting of the 1D subbands in this four-fold degenerate system, we have employed local band-gap engineering \cite{Kraft2018}, source-drain bias spectroscopy \cite{Patel1990,Patel1991,Martin-Moreno1992}, magnetoelectric subband-depopulation technique \cite{Vanwees1988a,Glazman1989}, and semi-phenomenological modelling. At lowest magnetic fields, clear steps of the QPC conductance quantization in units of $4\,e^2/h$ are observed. With increasing magnetic field, these steps split, forming a peculiar pattern combining steps of $e^2/h$, $2\,e^2/h$, and $4\,e^2/h$. Our model, based on the $2\times 2$ Hamiltonian~\cite{McCann2006,McCann2013}, agrees well with the full splitting of the Landau levels for the lowest two channels, as well as with the observed exotic merging/mixing of the $K$ and $K'$ valleys from pairs of 1D subbands with $(N,N+2)$ indices.\\    
	
For this study, we have used a device on which 1D confinement without edge currents was induced by local band-gap engineering and characterized by proximity-induced superconductivity and magneto-interferometry \cite{Kraft2018}. In those experiments, we used the displacement field created by the back and the split gate (BG and SG) voltage to locally open a band gap and confine the charge carriers in the QPC.
However, keeping this geometry does not allow us to drive the constriction to the low-density regime and observe the quantized conductance. In order to reach this regime, here we have added an overall top gate (TG) on an edge-connected BLG encapsulated between a bottom and top hexagonal boron nitride (hBN) multilayers, as depicted in Fig.\ref{fig:1} (see \cite{Kraft2018} and Supplemental Material (SM) \cite{SM} for details on the sample fabrication). As the BG counteracts and dominates over the SG for the control of the carrier density within the constriction, we use the TG to control the density not only by tuning the Fermi level \cite{Pyshkin2000,Hew2008,Hew2009} but also by opening a band gap in the 2D reservoirs and the constriction via the displacement field induced by BG and TG. Therefore, while keeping BG and SG constant, sweeping the TG voltage tunes the Fermi level, the confinement, and the band structure in the induced 1D system, down to full pinch-off \cite{SM}. A small perpendicular magnetic field $B=20\,$mT was applied to keep the Al leads in the normal metal state.

	\begin{figure}
		\centering
		\includegraphics{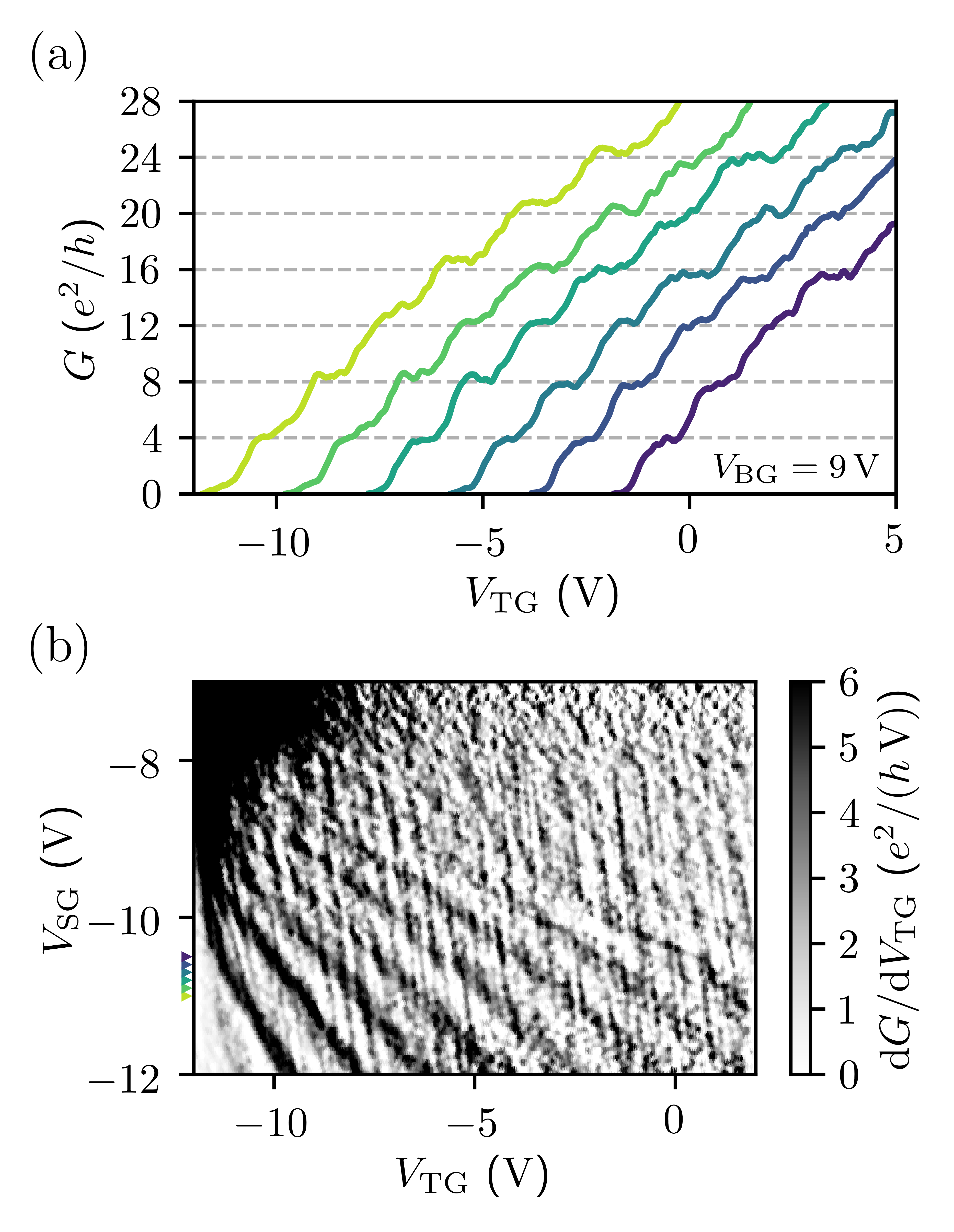}
		\caption{(a) Differential conductance $G$ as a function of TG voltage $V_\mathrm{TG}$ for different SG voltages $V_\mathrm{SG}$ from $-11.0\,$V (left) to $-10.5\,$V (right) with an increment of $0.1\,$V and at a constant BG voltage $V_\mathrm{BG}=9\,$V. The curves are shifted for clarity by $2\,$V between consecutive traces (the leftmost curve is not shifted). Well-quantized plateaus are observed in steps of $4\,e^2/h$. (b) Greyscale map of $\mathrm{d}G/\mathrm{d}V_\mathrm{TG}$ as a function of $V_\mathrm{TG}$ and $V_\mathrm{SG}$ at $V_\mathrm{BG}=9\,$V. Small markers denote the position of line cuts shown in (a).
		\vspace*{-0.5cm}}
		\label{fig:2}
	\end{figure}

In Fig.\,\ref{fig:2}a, the differential conductance $G$ through the QPC as a function of the TG voltage $V_\mathrm{TG}$ 
is displayed for different SG voltages $V_\mathrm{SG}$ at a constant BG voltage $V_\mathrm{BG} = 9$~V. The conductance curves are shifted for clarity and are based on raw data with no series resistance subtracted \cite{note}. Robust and stable quantized staircase in the conductance is observed with plateaus at integer values of $4\,e^2/h$ (see SM \cite{SM} for more details on the stability of the plateaus). We note that quantization of conductance appears only in a limited range of SG voltage $V_\mathrm{SG}$ for a given BG voltage $V_\mathrm{BG}$, when the Fermi level underneath the SG is placed in the induced band gap. 
In Fig.\,\ref{fig:2}b, a greyscale map of the differentiated differential conductance 
$\mathrm{d}G/\mathrm{d}V_\mathrm{TG}$ as a function of both $V_\mathrm{TG}$ and $V_\mathrm{SG}$ over an extended range of $V_\mathrm{SG}$ is displayed. The small colored triangles mark the SG values of the corresponding conductance traces shown in Fig.\,\ref{fig:2}a. The respective quantized plateaus are visible as large stripes that are tuned by both TG and SG. The plateaus, white in the greyscale map, are spreading with increasing $V_\mathrm{SG}$ that corresponds to an increasing subband level spacing as the confinement strengthened. The continuous evolution of the plateaus highlights the stability of the electrostatic confinement.

	\begin{figure}
		\centering
		\includegraphics[width=0.45\textwidth]{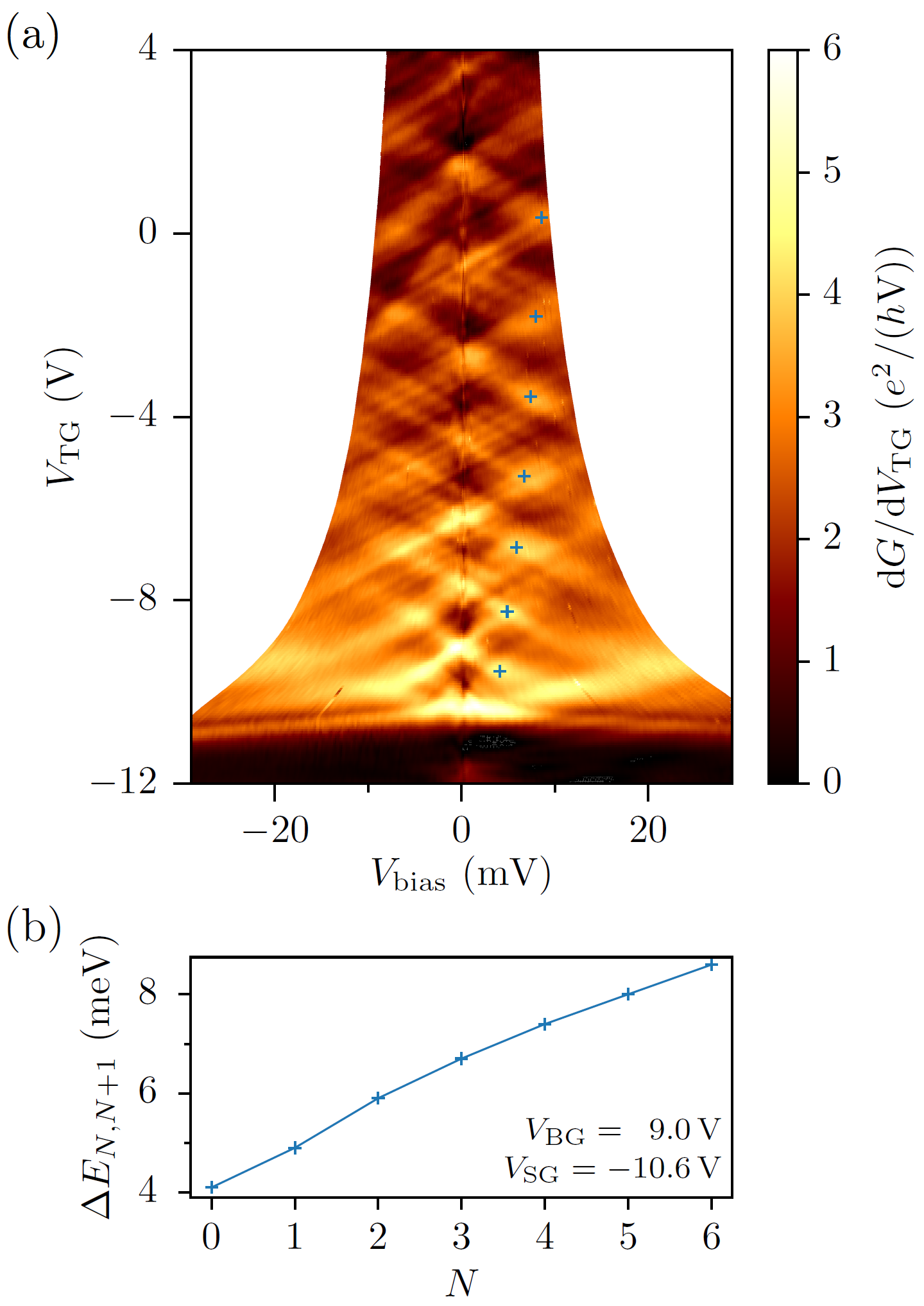}
		\caption{(a) Transconductance versus source-drain bias voltage $V_\mathrm{bias}$ and TG voltage $V_\mathrm{TG}$. Minima in $\mathrm{d}G/\mathrm{d}V_\mathrm{TG}$ correspond to plateaus in the $G(V_\mathrm{TG})$ curves. The resulting checkerboard pattern reveals an increasing energy level spacing with increasing subband index. Effect of the Fabry-P\'{e}rot interferences is clearly visible as lines parallel to the 1D subband dispersion lines. Blue crosses highlight the subband edge crossings representing the energy spacing $\Delta E_{N,N+1}$ between two consecutive 1D subbands of the QPC. In panel (b) $\Delta E_{N,N+1}$ shows a linear dependence as a function of the 1D subband indices $N$.
			\vspace*{-0.5cm}}
		\label{fig:3}
	\end{figure} 

It is important to note that no signs of anomalous features below the first quantized plateaus, namely the 0.7 structures \cite{Thomas1996,Micolich2011}, can be seen at the very low temperature of the experiment, $T\sim 20$~mK. One can also note that, within the plateaus in Fig.\,\ref{fig:2}a, additional fainted oscillations are observed. Superimposed on the oblique large stripes corresponding to the quantized plateaus, the additional oscillations appear as more fainted vertical lines in Fig.\,\ref{fig:2}b, mainly tuned by the TG but almost independent of the SG voltage. We attribute these conductance oscillations to Fabry-P\'{e}rot interferences arising from the two cavities formed by the contacts and the SG-induced barriers. We estimate the associated cavity size from the frequency of the resonances, yielding a length of about $230\,$nm which is in good agreement with the device geometry \cite{SM}. Strikingly, two phenomena that are both directly linked to the ballistic nature of the charge carrier transport but having two different physical origins, are visible concurrently. 

\begin{figure*}
		\centering
		\includegraphics[width=0.85\textwidth]{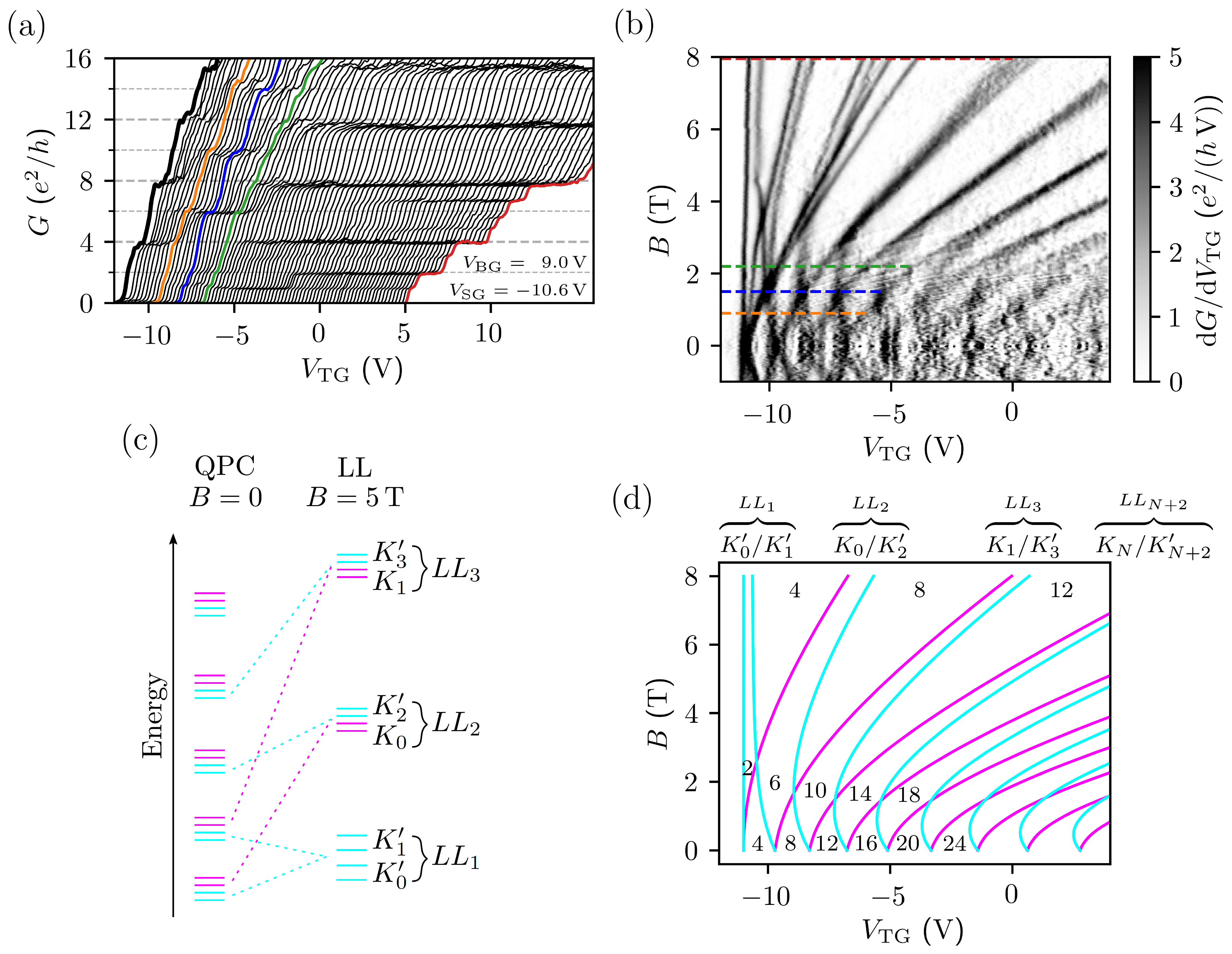}
		\caption{(a) Differential conductance $G$ as a function of TG voltage $V_\mathrm{TG}$ for different values of magnetic field  $B$ in steps of $100\,$mT at constant BG voltage $V_\mathrm{BG}=9\,$V and SG voltage $V_\mathrm{SG}=-10.6\,$V. The curves are shifted for clarity to the right by an offset of $2\,$V/T ($200\,$mV between consecutive curves). The thicker black line (not shifted) corresponds to the data acquired at $B=20\,$mT, which is shown in Fig.\,\ref{fig:2}. (b) Corresponding greyscale map of $\mathrm{d}G/\mathrm{d}V_\mathrm{TG}$ as a function of TG voltage $V_\mathrm{TG}$ and $B$. Colored dashed lines at $B=0.9\,$T (orange), $1.5\,$T (blue), $2.2\,$T (green) and $8.0\,$T (red) denote the linecuts associated with the highlighted conductance traces shown in panel (a). Transitions across magnetoelectric subbands appear as dark lines. (c) Energy level diagram of the QPC at zero and high magnetic field. (d) Valley subband dispersion in our QPC as a function of $B$ calculated with our model. The numbers displayed in the plot correspond to the quantized conductance values of the plateaus in units of $e^2/h$.}
		\label{fig:4}
	\end{figure*}

In order to characterize the 1D confinement of charge carriers and extract the subband spacing 
$\Delta E_{N,N+1}$, we have performed source-drain bias spectroscopy \cite{Patel1990,Patel1991,Martin-Moreno1992}. Fig.\,\ref{fig:3}a shows the colored map of the transconductance $\mathrm{d}G/\mathrm{d}V_{\mathrm{TG}}$ as a function of $V_{\mathrm{bias}}$ and $V_{\mathrm{TG}}$. Here, the plateaus appear in black, while colored lines represent transitions between the plateaus, \textit{i.e.} the subband edges. Subband edge crossings are marked by small crosses and $\Delta E_{N,N+1}$ increases approximatively linear from about 4 to 9\,meV for the first to the eighth subband. We note that this differs significantly from what is usually observed in QPCs, where one can easily model the system by a parabolic potential with $\Delta E_{N,N+1}$ increasing in the reversed fashion as the confinement is strengthened for lower subbands. Our system turns out to be more complex as the displacement field generated by the TG tunes the band structure within the 1D constriction. This makes the confinement in our QPC very challenging to model, which is beyond the scope of this work. In addition, we observe sets of lines parallel to the subband edge lines which can be attributed to the Fabry-P\'{e}rot interferences as aforementioned.    

To further analyze our QPC, we have studied the evolution of the 1D subband edges under a magnetic field $B$ perpendicular to the BLG plane. Figure\,\ref{fig:4}a shows $G$ as a function of TG voltage $V_\mathrm{TG}$ for different $B$ from 20\,mT (black thick curve) to 8\,T (red curve), from left to right in steps of $100\,$mT, at $V_\mathrm{BG}=9\,$V and $V_\mathrm{SG}=-10.6\,$V.  The curves are shifted for clarity by an offset of $200\,$mV between consecutive curves. A clear change in the quantization of the conductance steps is observed as the $B$ increases, from $4\,e^2/h$ to $e^2/h$ suggesting full lifting of the 1D subband degeneracy at high $B$. We note that the full splitting of the 1D subbands is fully ambipolar, therefore it occurs for both holes and electrons \cite{SM}. While the full lifting of the degeneracy has been observed in the quantum Hall regime in SLG \cite{Zimmermann2017} and BLG \cite{Overweg2018}, the transition from full degeneracy to full splitting has not been studied, to our knowledge. Figure\,\ref{fig:4}b displays the transconductance as a function of $B$ and $V_{\mathrm{TG}}$ of the data set of Fig.\,\ref{fig:4}a. This allows us to follow the complex 1D subband edge splitting of our QPC. Clear splitting of the 1D subbands, seen as dark lines in the greyscale map (bright parts represent quantized plateaus), is observed for the two first subbands (four lines each). However, splitting appears to be different at high $B$ for the higher subbands. The combination of quantizing electric and magnetic fields results into a complex splitting and bunching of the so-called magnetoelectric subbands \cite{Vanwees1988a}. 

In order to understand deeper the complex subband splitting on a qualitative level, 
we have developed a semi-phenomenological model \cite{SM} derived from the $2\times 2$ Hamiltonian of BLG \cite{McCann2006}. 
Since the band gap in the constriction is not too large (satisfying $\Delta\ll \gamma_1$, where $\gamma_1$ is the strongest interlayer hopping matrix element between A2-B1 atoms of graphene sheets), we ignore, for simplicity, the modification of the spectrum near the bottom of the conductance band and the top of the valence band (mexican-hat and trigonal-warping features; for the analysis of their effect on the QPC conductance, see Ref.~\cite{Knothe2018}).  We have defined our QPC as a BLG strip of width $W$.  With increasing magnetic field, the evolution of the eigenenergies and eigenstates for the $K$ and $K'$ valleys (neglecting the spin splitting) can be expressed as follows:
\begin{widetext}
\begin{align}
E_N^{K} &= \sqrt{\Delta^2 + (E_N^0)^2} \xrightarrow[B \rightarrow \infty]{} \sqrt{\Delta^2 + \omega_B^2 (N+1)(N+2)}, 
\qquad
&\Psi_{K} = 
\left(
\begin{matrix}
\varphi_N \\[0.1cm]
\dfrac{\hat{p}_+^2}{2m(E+\Delta)} \varphi_N
\end{matrix}
\right) 
\xrightarrow[B \rightarrow \infty]{}
\left(
\begin{matrix}
\tilde{\varphi}_N \\
\tilde{\varphi}_{N+2}
\end{matrix}
\right), 
\label{eq1}
\\
E_N^{K'} &= \sqrt{\Delta^2 + (E_N^0)^2} \xrightarrow[B \rightarrow \infty]{} \sqrt{\Delta^2 + \omega_B^2 (N-1)N},
\qquad
&\Psi_{K'} = 
\left(
\begin{matrix}
\varphi_N \\[0.1cm]
\dfrac{\hat{p}_-^2}{2m(E+\Delta)} \varphi_N
\end{matrix}
\right) 
\xrightarrow[B \rightarrow \infty]{}
\left(
\begin{matrix}
\tilde{\varphi}_N \\
\tilde{\varphi}_{N-2}
\end{matrix}
\right). 
\label{eq2}
\end{align}
\end{widetext}
Here $E_N^0$ denotes the size-quantization levels in the QPC at $B=0$, and the magnetic field,
characterized by the cyclotron frequency $\omega_B$, is included through the shift in momentum operators $\hat{p}_{\pm} = \hat{p}_x - e A_x/c \pm ( i \hat{p}_y - i e A_y/c)$ by the corresponding vector potential.

At $B = 0$, the energy levels are degenerate in $K$ and $K'$ valleys. The components of the spinors are  given by the electron wave-functions in a 1D quantum well: $\varphi_N$ ($N = 0, 1, 2, ...$). With increasing $B$, the size quantization wave-function trends to a harmonic-oscillator wave-function with the same number $\varphi_N \xrightarrow[B \rightarrow \infty]{} \tilde{\varphi}_{N}$. This results in degenerate Landau levels in strong $B$ for the valleys $K$ and $K'$ coming from two different subbands with indices $N$ differing by 2, as shown in Eqs.\,(\ref{eq1}) and (\ref{eq2}). Figure\,\ref{fig:4}c depicts schematically the pattern of energy levels in the QPC at zero and at high $B$, while Fig.\,\ref{fig:4}d shows the evolution of the 1D subbands with magnetic field resulting from Eqs.\,(\ref{eq1}) and (\ref{eq2}). Comparing this plot with Fig.\,\ref{fig:4}b, we see that our simplified model captures
the main qualitative features of the valley splitting induced by magnetic field. An additional splitting 
of Landau levels $LL_1$ and $LL_2$ observed in the experiment can be attributed to the renormalization (most prominent at the lowest densities) of the Zeeman splitting (neglected in our model) by the Coulomb interaction. 

Finally, although we focused on the most clear conductance quantization in steps of $4\,e^2/h$ characteristic of a strong constriction, we mention that at smaller split-gate voltage, $V_\text{SG}<-9.5\,$V at BG voltage $V_\mathrm{BG}=9\,$V, we observe a vanishing of the first plateau and a new $8\,e^2/h$-step in the quantization of the lowest subband appears (see Fig.\,2b). The additional degeneracy is also apparent in the depopulation of the magnetoelectric subbands (see \cite{SM}). This is in agreement with the prediction of Ref.\,\cite{Knothe2018} about the possibility of ``accidental'' degeneracy of the size-quantized subbands in smoother constrictions that results from the mexican-hat feature of the spectrum with relatively large gap.

To conclude, we have studied the valley splitting in a BLG QPC subject to magnetic field. We have measured the quantized conductance  through the QPC and observed robust and stable conductance steps quantized in units of $4\,e^2/h$, as expected for this four-fold degenerated system with a small band gap.  Using source-drain bias spectroscopy, we have determined the 1D subband spacing $\Delta E_{N,N+1}$ which reveals an apparent unconventional confinement. Under high magnetic field $B$ perpendicular to the sample plane, both spin and valley degeneracy fully lift as the density is lowered, \textit{i.e.} as both confinement and Coulomb interactions are enhanced, magnetoelectric subbands are formed \cite{Vanwees1988a} reflecting the peculiar pseudospin structure of BLG. Our semi-phenomenological model demonstrates that the QPC size-quantized modes undergo subband mixing and merging of the $K$ and $K'$ valleys with non-consecutive indices. Indeed, for higher modes, the conductance quantization in units of $4\,e^2/h$ is restored in strong magnetic fields. At the same time, for the lowest two resulting Landau levels, the Zeeman splitting is enhanced by interactions, leading to the observed steps of $e^2/h$ in the conductance (red curve in Fig.\,\ref{fig:4}a). At intermediate fields, a complex pattern of the energy levels produces also the conductance steps of $2\,e^2/h$ due to valley splitting (orange and green curve in Fig.\,\ref{fig:4}a), as well as the restored but shifted sequence $(N+1/2)\cdot 4\,e^2/h$ when splitted lines from neighboring subbands are crossing (blue curve in Fig.\,\ref{fig:4}a). Our study thus demonstrates high versatility of band engineering in BLG and provides an input for developing graphene-based valleytronics.

This work was partly supported by Helmholtz society through program STN, 
the Russian Science Foundation (I.V.K., A.P.D and I.V.G., Grant No. 17-12-01182, theoretical modelling),
the Foundation for the advancement of theoretical physics and mathematics BASIS (I.V.K.), 
and the DFG via the project DA 1280/3-1 and the FLAG-ERA JTC2017 Project GRANSPORT (GO  1405/5-1,
Karlsruhe node). 

\textit{Note added:} When submitting the manuscript we became aware of the preprint arXiv:1809.01920 which reported on the conductance quantization in a similar structure but with a top gate covering only the QPC region.

\end{document}


\title{Supplemental Material:\\ 
"Valley subband splitting in bilayer graphene quantum point contact"}

\author{R. Kraft}
\affiliation{Institute of Nanotechnology, Karlsruhe Institute of Technology, D-76021 Karlsruhe, Germany}

\author{I.V. Krainov}
\affiliation{A.F. Ioffe Physico-Technical Institute, 194021 St. Petersburg, Russia}

\author{V. Gall}
\affiliation{Institute of Nanotechnology, Karlsruhe Institute of Technology, D-76021 Karlsruhe, Germany}
\affiliation{Institute for Condensed Matter Theory, Karlsruhe Institute of Technology, D-76128 Karlsruhe, Germany}

\author{A.P. Dmitriev}
\affiliation{A.F. Ioffe Physico-Technical Institute, 194021 St. Petersburg, Russia}

\author{R. Krupke}
\affiliation{Institute of Nanotechnology, Karlsruhe Institute of Technology, D-76021 Karlsruhe, Germany}
\affiliation{Department of Materials and Earth Sciences, Technical University Darmstadt, Darmstadt, Germany}

\author{I.V. Gornyi}
\affiliation{Institute of Nanotechnology, Karlsruhe Institute of Technology, D-76021 Karlsruhe, Germany}
\affiliation{A.F. Ioffe Physico-Technical Institute, 194021 St. Petersburg, Russia}
\affiliation{Institute for Condensed Matter Theory, Karlsruhe Institute of Technology, D-76128 Karlsruhe, Germany}

\author{R. Danneau}
\affiliation{Institute of Nanotechnology, Karlsruhe Institute of Technology, D-76021 Karlsruhe, Germany}

\maketitle	

\section{Sample fabrication and experimental condition details}

For this study, we have used a bilayer graphene (BLG) device presented in \cite{Kraft2018} and added an overall top gate (TG). The original device is an edge-connected hBN-BLG-hBN heterostructure \cite{Wang2013} (top and bottom hBN of about $38\,$nm and $35\,$nm thick respectively) placed onto a pre-patterned back gate (BG) designed on a sapphire substrate, covered by an additional dielectric layer Al$_2$O$_3$ ($20\,$nm) deposited by atomic layer deposition (ALD). The BLG is edge-connected \cite{Wang2013} with Ti/Al electrodes and the QPC is defined by a split gate (SG) designed on top of the heterostructure (see \cite{Kraft2018} for details). The entire sample was then covered by an extra layer of Al$_2$O$_3$ ($30\,$nm) deposited by ALD and an overall Ti/Cu  top gate (TG).

The electrical measurements have been performed in a $^3$He/$^4$He dilution refrigerator BF-LD250 from BlueFors at a base temperature below 20mK unless otherwise mentioned. The sample is probed in a two-terminal configuration using standard low-frequency ($\sim 13\,$Hz) lock-in technique with an AC excitation ranging from 1 to 20\,$\mu$V. Finally, all magnetic fields used in these experiments were applied perpendicular to the BLG plane.

\section{Effect of split and top gate}
	
	\begin{figure*}
		\centering
		\includegraphics{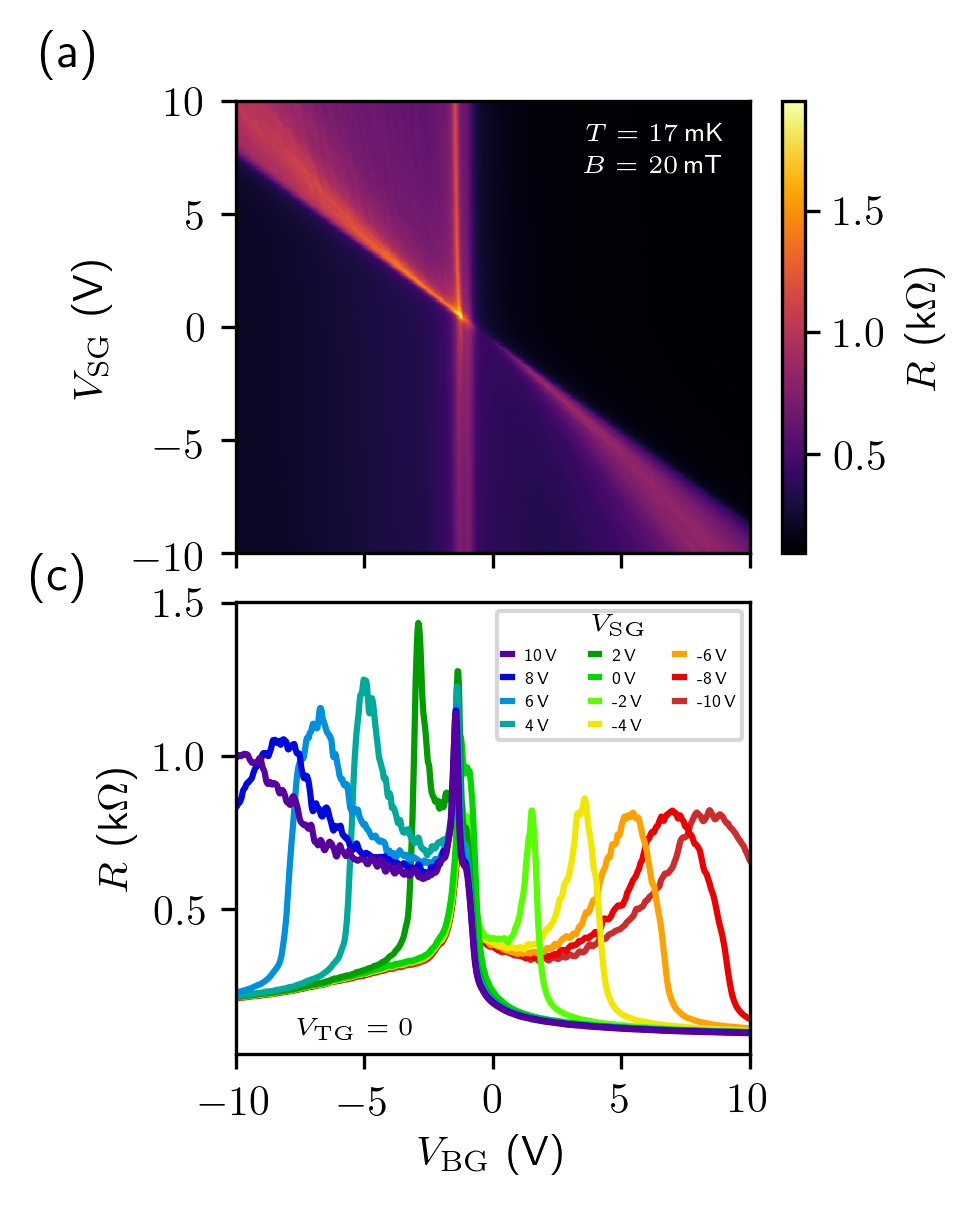}
		\includegraphics{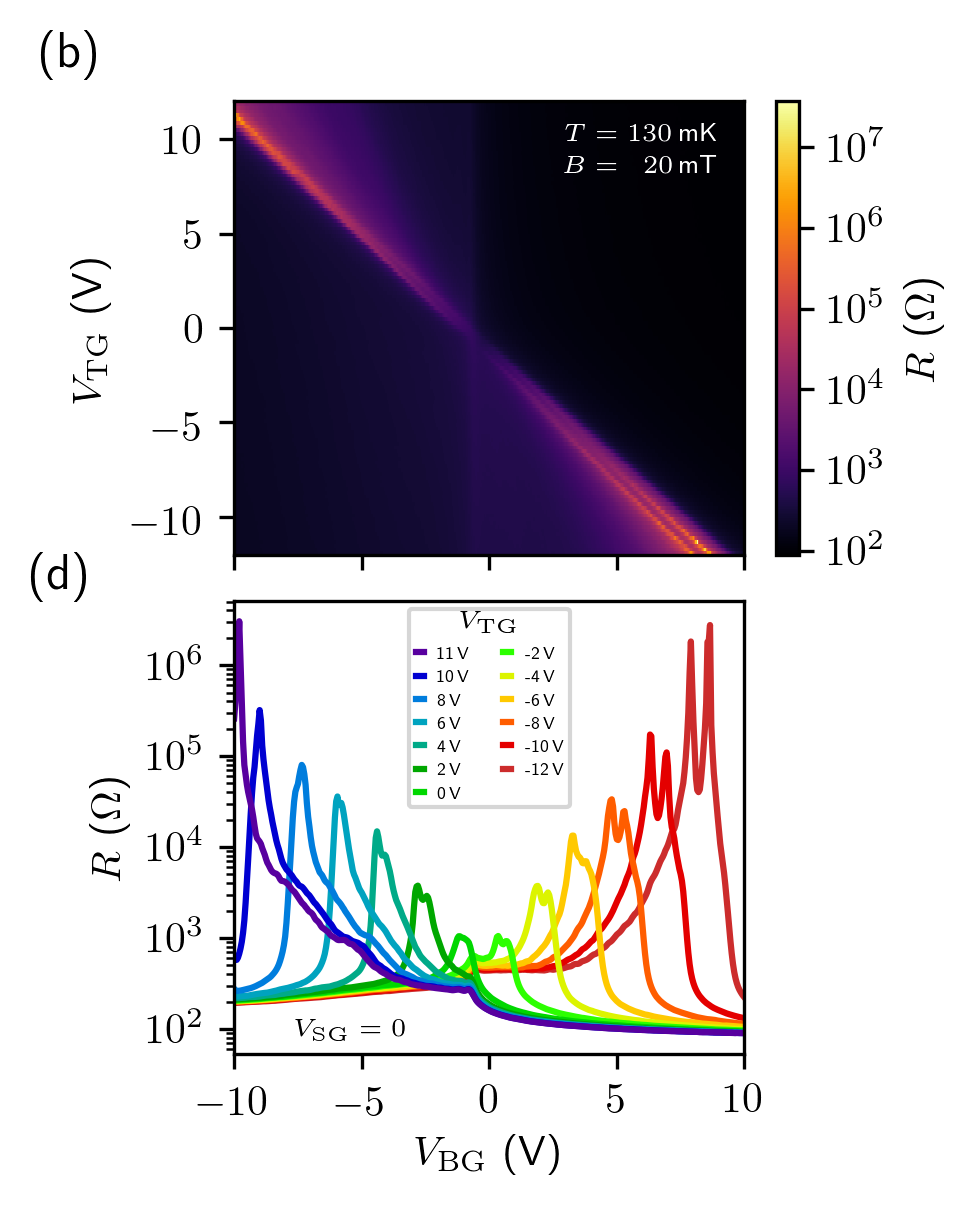}
		\caption{(a), (b) Color maps of the differential resistance $R$ as a function of BG voltage $V_\mathrm{BG}$ and SG voltage $V_\mathrm{SG}$ (a) or TG voltage $V_\mathrm{TG}$ (b). The diagonal high resistance line corresponds to the condition of charge neutrality in the split-gated or top-gated region respectively, while the vertical line only tuned by the BG corresponds to charge neutrality in the regions that are independent of either SG or TG respectively. The displacement field $D$ increases along the diagonal line, resulting in the opening of a band gap while the Fermi level is kept constant at zero energy in the band gap center. (c), (d) Linecuts of the maps showing the differential resistance $R$ as a function of BG voltage $V_\mathrm{BG}$ for various SG voltages (c) or TG voltages (d).}
		\label{fig:S1}
	\end{figure*}
	
Here we analyze the effect on transport of the SG and TG in the presence of a BG voltage $V_\mathrm{BG}$. The differential resistance $R$ of the device is shown in Fig.\,\ref{fig:S1}a,\,b as a function of BG voltage $V_\mathrm{BG}$ and either SG voltage $V_\mathrm{SG}$ or TG voltage $V_\mathrm{TG}$. During each measurement, the unused gate was kept grounded. Corresponding horizontal linecuts are presented in Fig.\,\ref{fig:S1}c,\,d.
	
Two lines of resistance maxima are visible in the color maps. In case of the SG, the vertical line corresponds to charge neutrality of the reservoirs while the diagonal line corresponds to charge neutrality of the SG region which is tuned by both BG and SG. As the gap develops with increasing displacement field $D$ along the diagonal resistance line, conductance is fully suppressed underneath the SG. However, as the SG does not cover entirely the width of the device, the resistance maximum does not go beyond $2\,$k$\Omega$ due to the remaining conducting channel between the SG electrodes (see \cite{Kraft2018} for detailed explanations).

In contrast, the TG fully covers the device, \textit{i.e.} the entire width of the BLG layer. Then the diagonal resistance maximum rises up to resistance values in the order of $10\,$M$\Omega$, \textit{i.e.}, maximum limit of our lock-in detection technique (similarly as in \cite{Oostinga2007}). We note that a double maximum in the resistance along the displacement field line is observed, which might be explained by the two different top-gated regions of our sample, \textit{i.e.} reservoirs and QPC, where a partial screening by the SG may lead to slightly different capacitive coupling of the TG.

\section{Quantized conductance and magnetic field dependence under different gate conditions}

\begin{figure}
		\centering
		\includegraphics{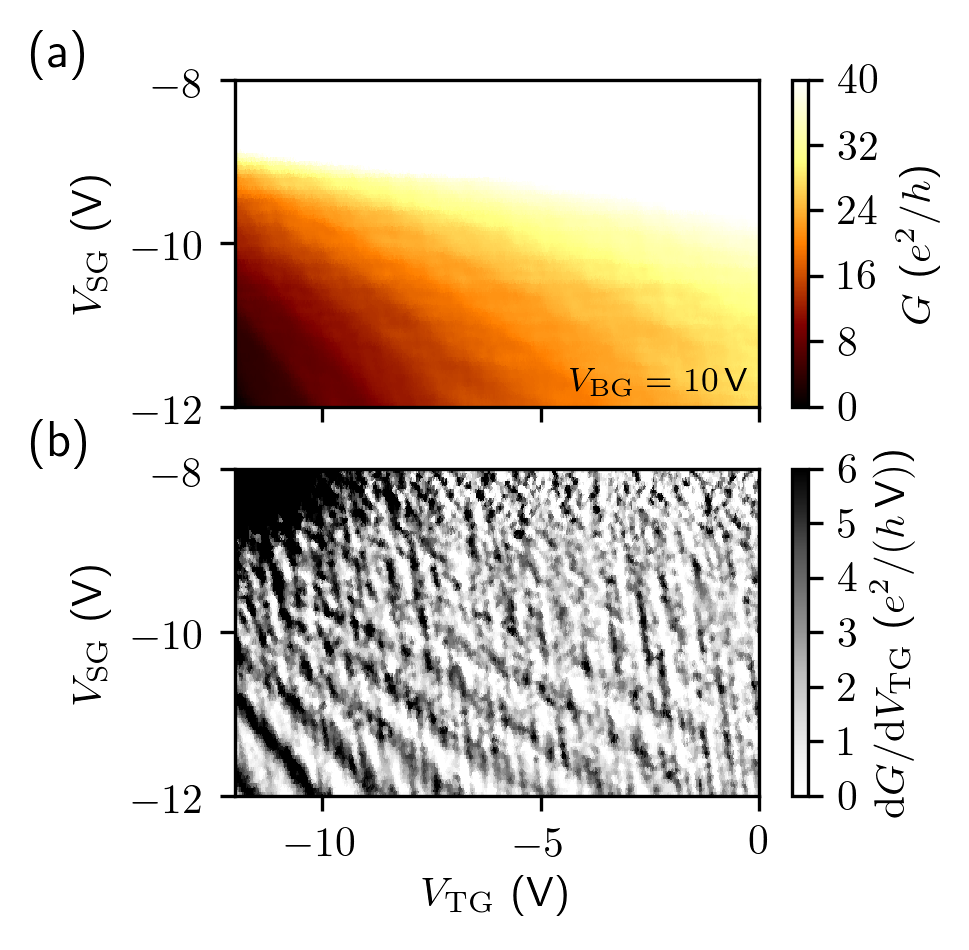}
		\caption{(a) Differential conductance $G$ and (b) differentiated differential conductance $\mathrm{d}G/\mathrm{d}V_\mathrm{TG}$ as functions of TG voltage $V_\mathrm{TG}$ and SG voltage $V_\mathrm{SG}$ at constant BG voltage $V_\mathrm{BG}=10\,$V.}
		\label{fig:S2}
	\end{figure}

\begin{figure}
		\centering
      \includegraphics{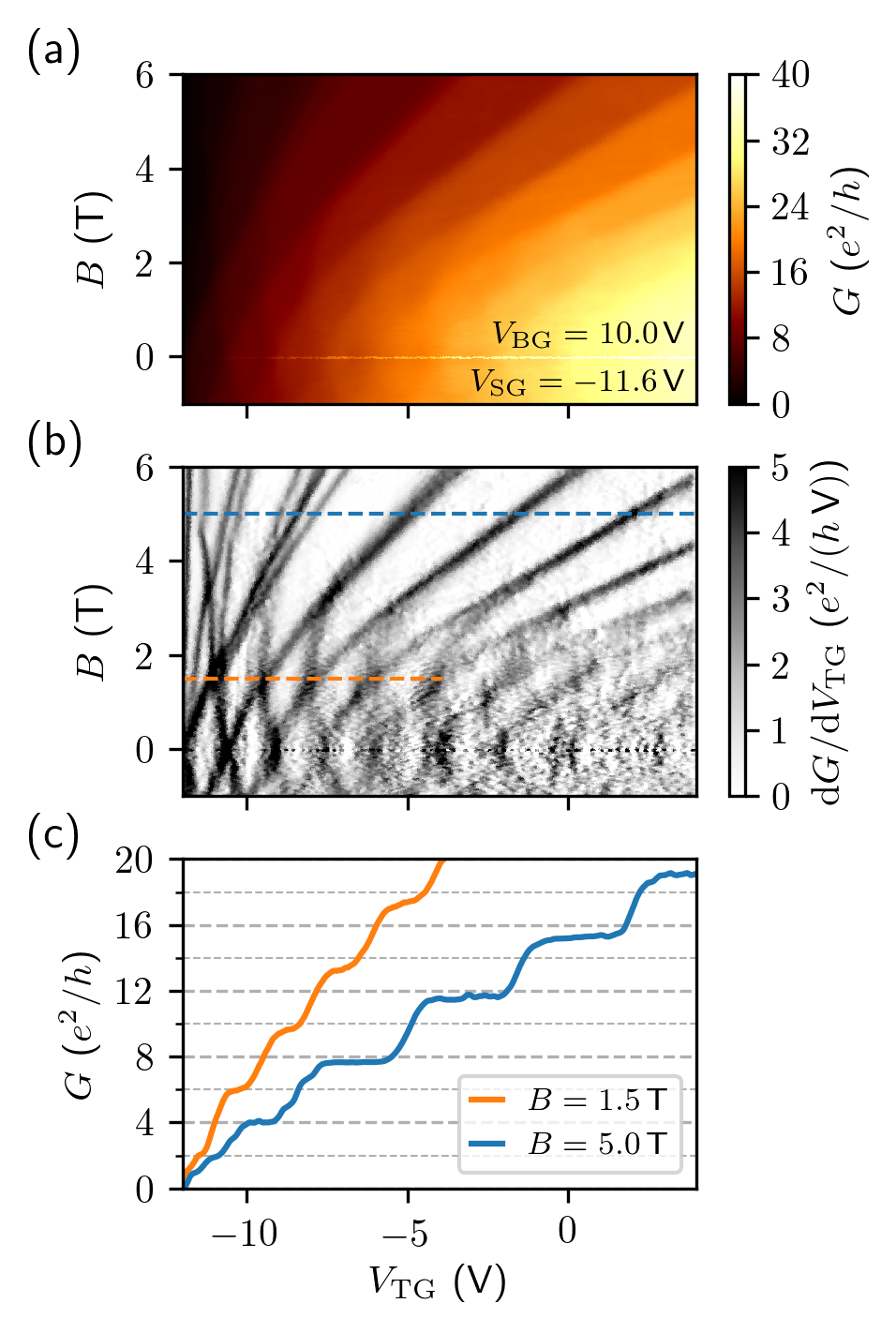}
      \caption{(a) Differential conductance $G$ and (b) differentiated differential conductance $\mathrm{d}G/\mathrm{d}V_\mathrm{TG}$ as functions of TG voltage $V_\mathrm{TG}$ and magnetic field $B$ at constant BG voltage $V_\mathrm{BG}=10\,$V and SG voltage $V_\mathrm{SG}=-11.6\,$V. The first subband is reached just at the limit of the maximum applied TG voltage $V_\mathrm{TG}=-12\,$V. (c) Differential conductance $G$ as a function of TG voltage $V_\mathrm{TG}$ for different $B$. Curves correspond to linecuts of the above panels as marked by the dashed lines in (b).}
      \label{fig:S3}
	\end{figure}

We have measured our QPC under various gate conditions. Figure\,\ref{fig:S2} shows the differential conductance $G$ and differentiated differential conductance $\mathrm{d}G/\mathrm{d}V_\mathrm{TG}$ as functions of TG voltage $V_\mathrm{TG}$ and SG voltage $V_\mathrm{SG}$ at BG voltage $V_\mathrm{BG}=10\,$V. While the confinement is changed with respect to the data presented in the main text, the main features are conserved. Quantized conductance plateaus are observed as large stripes, whereas superimposed vertical conductance oscillations are due to Fabry-P\'{e}rot resonances of the reservoirs. As shown in the main text, the magnetic depopulation \cite{Vanwees1988a} of the 1D subbands shows a complex pattern of splitting and bunching of magnetoelectric subbands (see Fig.\,\ref{fig:S3}).  
	  	
\section{Ambipolar QPC}

\begin{figure}
		\centering
		\includegraphics{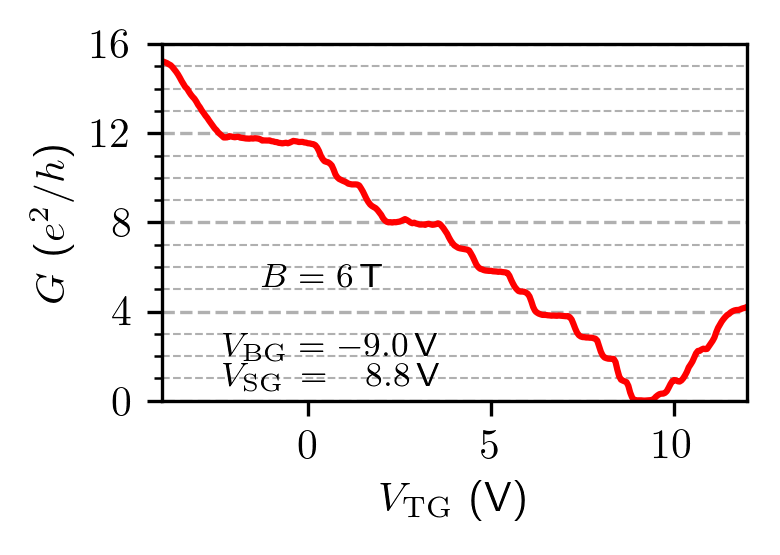}
		\caption{Differential conductance $G$ as a function of TG voltage $V_\mathrm{TG}$ at constant BG voltage $V_\mathrm{BG}=-9\,$V and SG voltage $V_\mathrm{SG}=8.8\,$V for $B=6\,$T.}
		\label{fig:S4}
	\end{figure}

We have probed our device in the opposite gate polarity, \textit{e.g.}, with BG voltage $V_\mathrm{BG}=-9.0\,$V and positive SG voltage $V_\mathrm{SG}=8.8\,$V. Therefore, while the gap opens, the Fermi level is positioned in the valence band in the 1D constriction and the 2D reservoirs. Figure\,\ref{fig:S4} displays the differential conductance $G$ as a function of TG voltage $V_\mathrm{TG}$ at $B=6\,$T. By applying an increasing positive TG voltage $V_\mathrm{TG}$ the hole subbands are depopulated, resulting in a stepwise decrease of the conductance. The spin- and valley-degeneracy of the magnetoelectric subbands is fully lifted below $G=12\,e^2/h$ and plateaus appear in steps of $e^2/h$. When the Fermi level is finally tuned into the gap, the conduction through the channel is pinched-off. However, further increase of the TG voltage $V_\mathrm{TG}$ results in tuning the Fermi level into the conduction band and thus the population of the system with electrons. Then, we observe a stepwise increase of the conductance with a full splitting of the Landau levels in steps of $e^2/h$ below the first plateau. This highlights the ambipolarity of our system where the splitting of the Landau levels in steps of $e^2/h$ can be probed continuously for both types of charge carriers.  

	\begin{figure}
		\centering
		\includegraphics{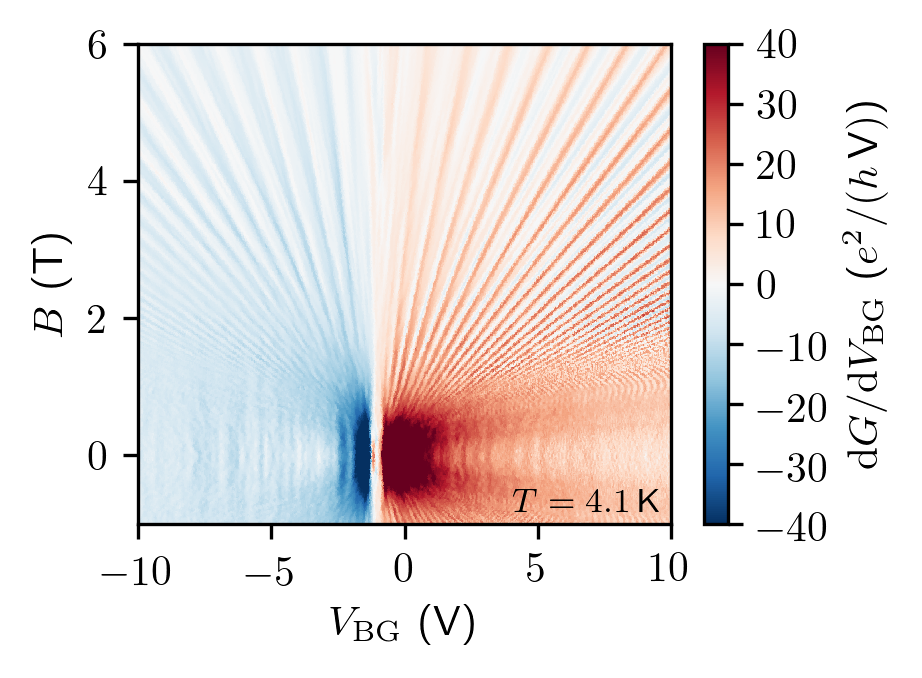}
		\caption{Differentiated conductance $\mathrm{d}G/\mathrm{d}V_\mathrm{BG}$ as a function of the BG voltage $V_\mathrm{BG}$ and magnetic field $B$ at SG and TG voltage $V_\mathrm{SG/TG}=0$, measured at a temperature $T=4.1\,$K. The gate configuration resembles the scenario of the 2D device without constriction, though it is not measured exactly at charge neutrality of SG and TG because of small residual doping due to slightly shifted Fermi levels.}
		\label{fig:S5}
	\end{figure}

\section{Landau level fan of the 2D system}

The magnetic field dependence at zero SG and TG voltage of the BLG is displayed in Fig.\,\ref{fig:S5}.  Under these conditions the differentiated differential conductance $\mathrm{d}G/\mathrm{d}V_\mathrm{TG}$ as a function of BG voltage $V_\mathrm{BG}$ and magnetic field $B$ shows a regular Landau level fan diagram. We note that  no valley splitting can be seen.

\section{Fabry-P\'{e}rot interferences}
	 
 	\begin{figure*}
		\centering
		\includegraphics{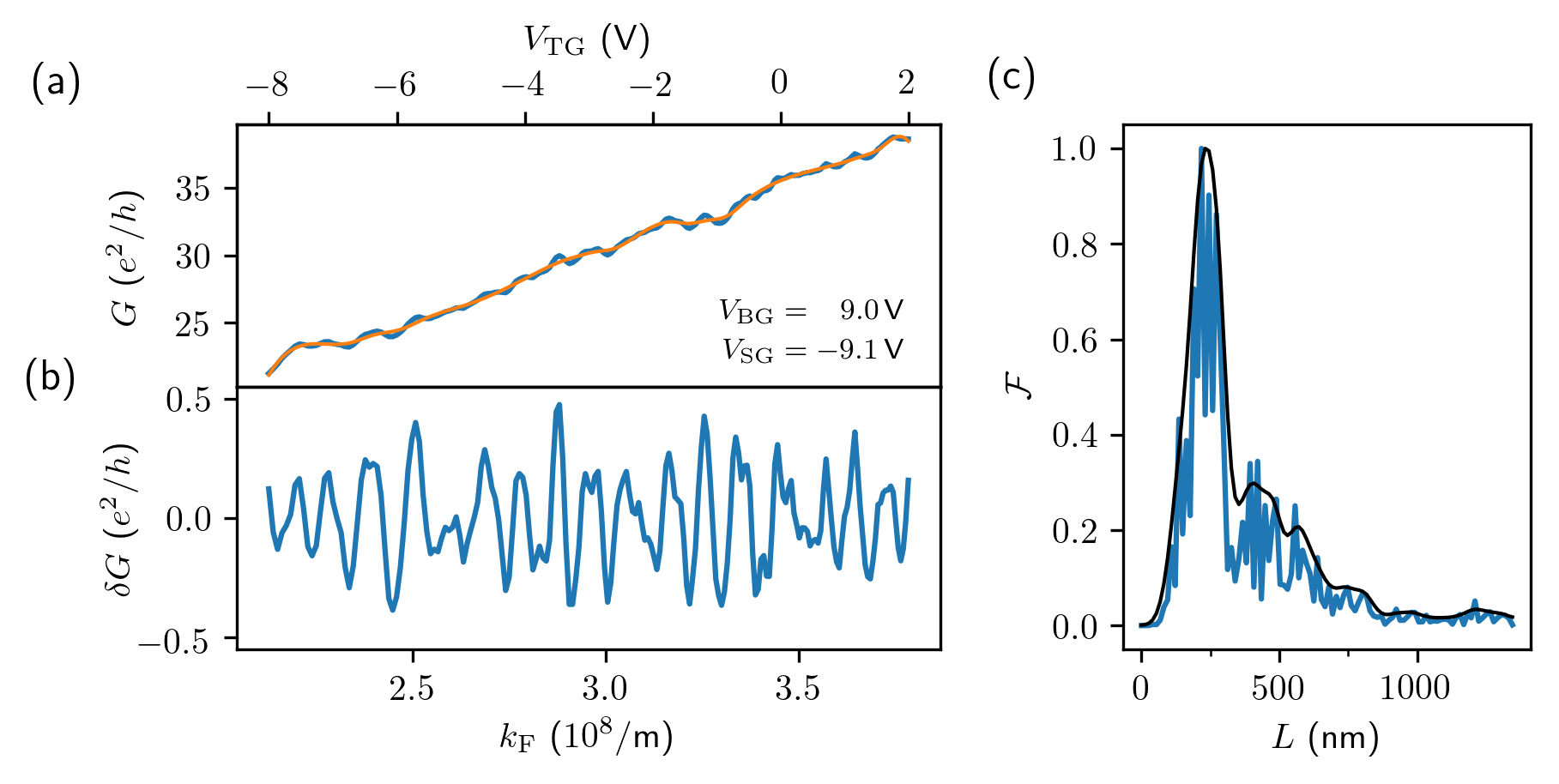}
		\caption{(a) Differential conductance $G$ and (b) oscillating part of the conductance $\delta G$ as functions of the TG voltage $V_\mathrm{TG}$ (upper abscissa) and Fermi wave vector $k_\mathrm{F}$ (lower abscissa) at constant BG voltage $V_\mathrm{BG}=9\,$V and SG voltage $V_\mathrm{SG}=-9.1\,$V. The oscillating part is obtained by subtracting the smooth background conductance (orange) from the raw conductance (blue). (c) Normalized Fourier transform of $\delta G$ as a function of length $L$, yielding the frequency spectrum of the oscillations corresponding to the size of the cavity at the resonance condition. The smooth curve (black) is obtained by convolving the more noisy raw signal (blue) with a gaussian filter.}
		\label{fig:S6}
	\end{figure*}
 
Here, we analyze the Fabry-P\'{e}rot interferences observed as conductance oscillations as aforementioned. As we see in Fig.\,2b of the main text and Fig.\,\ref{fig:S2} in this Supplemental Material (SM), the interferences are mainly tuned by the TG voltage. Therefore, the resonances may occur in a cavity formed by the non-splitgated part of the device.

We estimate the size of the cavity associated with the observed Fabry-P\'{e}rot interferences. Figure\,\ref{fig:S6}a shows the differential conductance $G$ as a function of TG voltage $V_\mathrm{TG}$ and Fermi wave vector $k_\mathrm{F}$ respectively (at $V_\mathrm{BG}=9.0\,$V and $V_\mathrm{SG}=-9.1\,$V), corresponding to Fig.\,2b of the main text. The oscillating part of the conductance $\delta G$, plotted in Fig.\,\ref{fig:S6}b, is obtained by subtracting the smooth background from the conductance. The size of the cavity is directly linked to the oscillation frequency at resonance condition $L=j\cdot\frac{\mathrm{\pi}}{k_\mathrm{F}}$, with $j$ an integer number and thus can be extracted by performing a Fourier transform. The resulting frequency spectrum is shown in Fig.\,\ref{fig:S6}c. A pronounced peak is observed at about $230\,$nm, being in good agreement with the physical distance between contacts and SG.\\

\section{Stability of the electrostatically induced QPC}
	
	\begin{figure}
		\centering
		\includegraphics{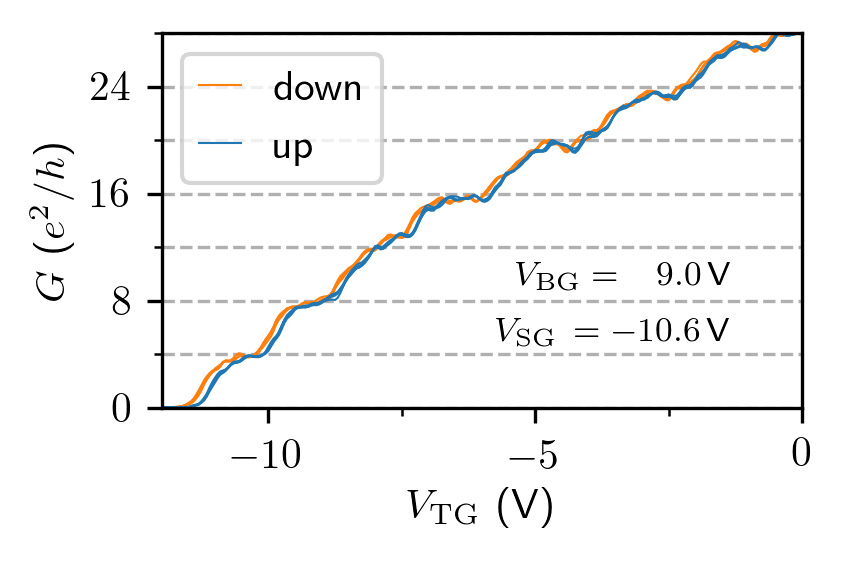}
		\caption{Differential conductance $G$ as a function of TG voltage $V_\mathrm{TG}$ at constant BG voltage $V_\mathrm{BG}=9.0\,$V and SG voltage $V_\mathrm{SG}=-10.6\,$V. The plot showing four curves each up and down for consecutive measurements.}
		\label{fig:S7}
	\end{figure}	 
	 
Here, we test the stability of the conductance quantization. Multiple TG sweeps (up and down) are recorded under the same confinement condition as presented data in the main text ($V_\mathrm{BG}=9.0\,$V and $V_\mathrm{SG}=-10.6$V). In total four curves each up ($V_\mathrm{TG}=-12\,$V$\rightarrow 0$) and down ($V_\mathrm{TG}=0\rightarrow -12\,$V) are measured and plotted in Fig.\,\ref{fig:S7}. We note that all four curves perfectly sit on top of each other making them indistinguishable and both features conductance plateaus and Fabry-P\'{e}rot interferences are fully reproduced. We note that a very small hysteresis between up- and down-sweeps is visible.

\section{Source-drain bias spectroscopy and 1D subband spacing}
	
	\begin{figure*}
		\centering
		\includegraphics[width=0.95\textwidth]{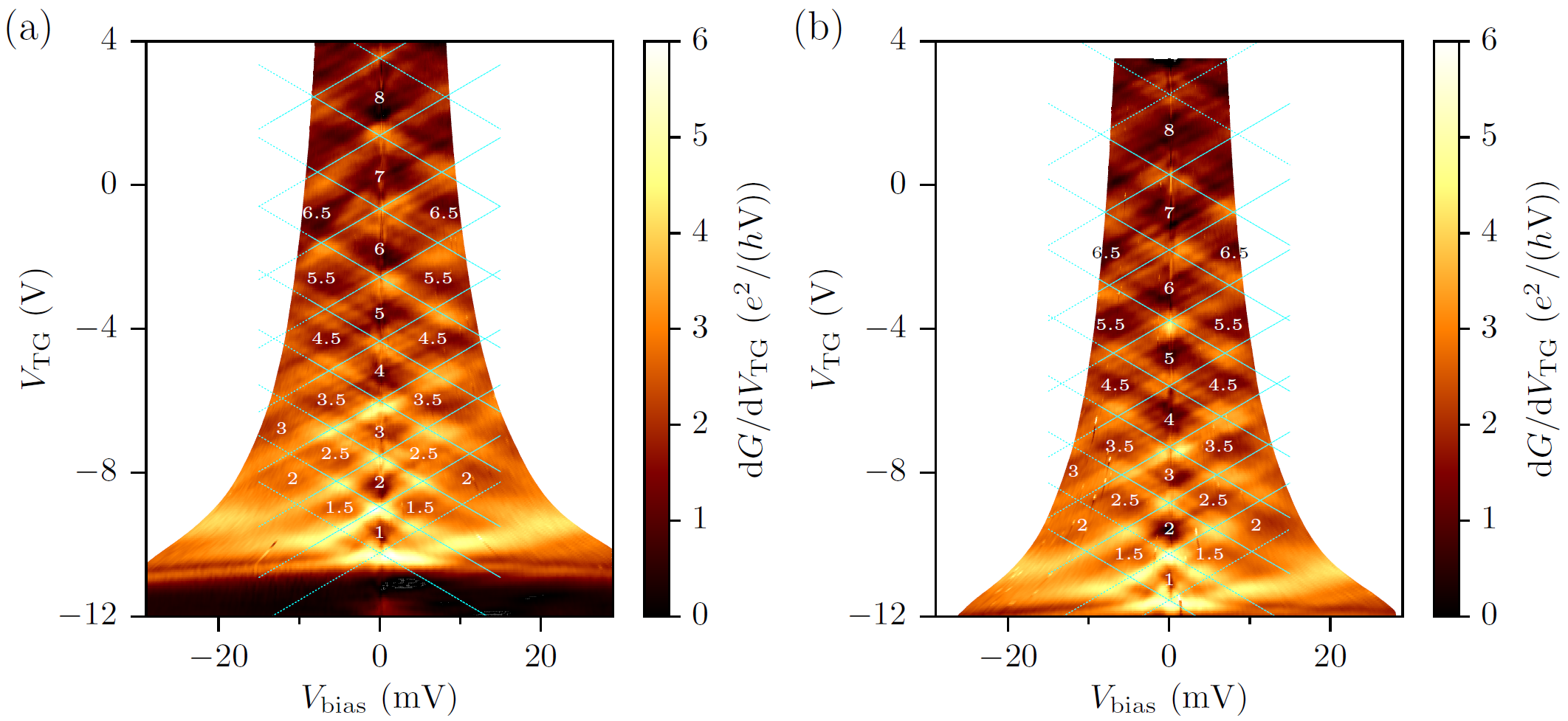}
		\caption{Transconductance as a function of source-drain bias voltage $V_\mathrm{bias}$ and TG voltage $V_\mathrm{TG}$ at (a) BG voltage $V_\mathrm{BG}=9.0\,$V and SG voltage $V_\mathrm{SG}=-10.6\,$V and (b) $V_\mathrm{BG}=10.0\,$V and $V_\mathrm{SG}=-11.6\,$V. The displayed numbers in the plots correspond to the quantized conductance value of the plateaus in units of $4\,e^2/h$ and cyan lines trace the transitions between plateaus.}
		\label{fig:S8}
	\end{figure*}

Source-drain bias spectroscopy is commonly used to probe the energy level spacing of the 1D subbands formed in QPCs \cite{Patel1990,Patel1991,Martin-Moreno1992}. We have used this measurement technique not only to extract the 1D subband spacing as described in the main text, but we also utilized it to extract the coupling factor $\alpha_\mathrm{TG}$. This parameter allows the conversion between $V_\mathrm{TG}$ that is applied to depopulate the 1D constriction and the energy levels of the QPC. $\alpha_\mathrm{TG}$ is then used to plot valley subband dispersion as a function of magnetic field obtained by our model.     	
	
In Fig.\,\ref{fig:S8} color maps of the transconductance as functions of source-drain bias voltage $V_\mathrm{bias}$ and TG voltage $V_\mathrm{TG}$ are shown under two similar confinement conditions: Figure\,\ref{fig:S8}a at BG voltage $V_\mathrm{BG}=9.0\,$V and SG voltage $V_\mathrm{SG}=-10.6\,$V (see Fig.\,3 of the main text) and Fig.\,\ref{fig:S8}b at BG voltage $V_\mathrm{BG}=10.0\,$V and SG voltage $V_\mathrm{SG}=-11.6\,$V (corresponding to a confinement condition as presented in Fig.\,3 of the SM). The quantized conductance plateaus are labeled with the associated conductance values in units of $4\,e^2/h$. In both cases, as shown in the main text, the subband spacing increases as the confinement is weakened as well as the effect of Fabry-P\'{e}rot interferences (see main text).    

Additionally, we plot an overlaying set of cyan lines, marking transitions across subband edges, which are described by the expression	${ \alpha_\mathrm{TG}e(V_\mathrm{TG}-V_\mathrm{TG}^0) = E_N^\mathrm{QPC} \pm eV_\mathrm{bias}/2 }$\,. To fit the energy levels $E_N^\mathrm{QPC}$ we considered a linearly increasing energy level spacing of the size-quantized subbands $ \Delta_{N,N+1} \propto N $ (see Fig.\,3b of the main text). The resulting set of lines are in qualitative good agreement with the energy levels of the subbands in the transconductance pattern.

From the slope of these lines we find the proportionality factor $\alpha_\mathrm{TG}=3.8\cdot 10^{-3}$ converting TG voltage $V_\mathrm{TG}$ into energy ${E=\alpha_\mathrm{TG}e(V_\mathrm{TG}-V_\mathrm{TG}^0)}$. The extracted gate coupling parameter from source-drain bias spectroscopy is used for plotting, together with the fitted energy levels, the energy levels of the magnetoelectic subbands derived from our model in terms of TG voltage $V_\mathrm{TG}$.    
	 
\section{Weak confinement regime}	 

	\begin{figure}
		\centering
		\includegraphics[width=0.41\textwidth]{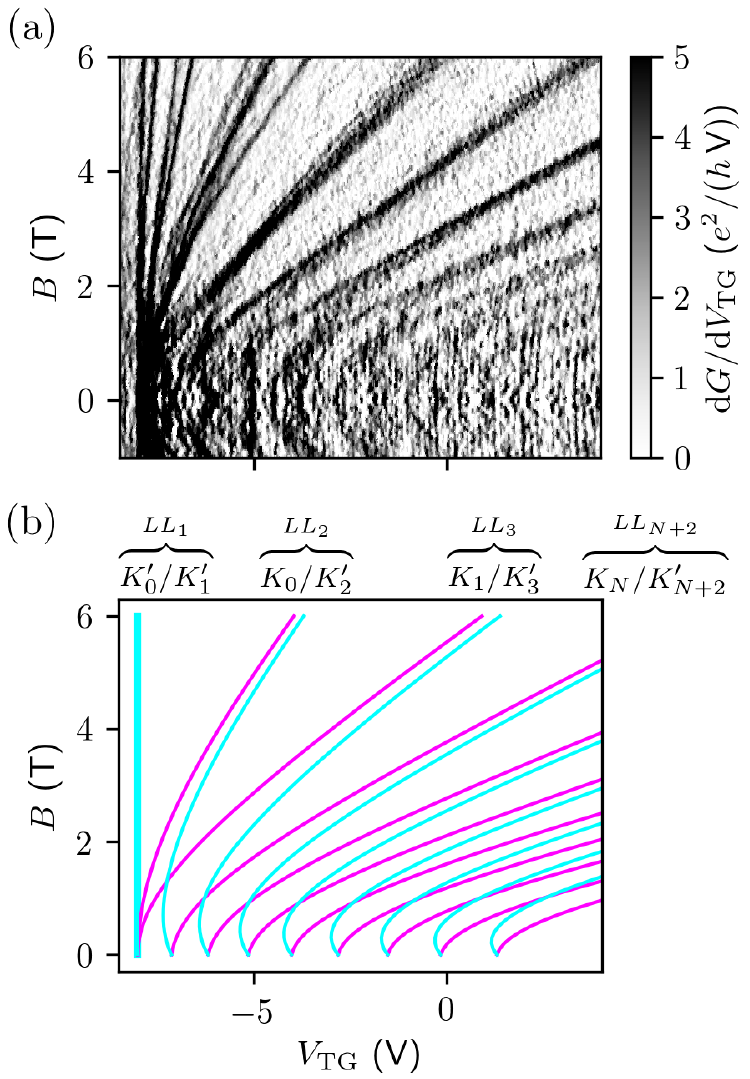}
		\caption{(a) Greyscale map of the differentiated differential conductance $\mathrm{d}G/\mathrm{d}V_\mathrm{TG}$ as a function of TG voltage $V_\mathrm{TG}$ and magnetic field $B$ at constant BG voltage $V_\mathrm{BG}=6.0\,$V and SG voltage $V_\mathrm{SG}=-6.7\,$V. Magnetoelectric subbands are visible as dark lines. (b) Valley subband dispersion as a function of magnetic field $B$ calculated with our model. Two colors distinguish between the two valleys. Lines here are understood as double lines, reflecting the twofold spin degeneracy in the model, except for the first thick cyan line , reflecting a fourfold (spin and ``touching-band'') degeneracy. In strong magnetic field, the Zeeman
		splitting (neglected in the model), enhanced at low densities by the Coulomb interaction, lifts the spin degeneracy. A similar effect lifts the ``accidental'' mexican-hat degeneracy.}
		\label{fig:S9}
	\end{figure}

Here we present data and modelling of the 1D subband splitting in a weak confinement regime as defined in \cite{Knothe2018}. An additional ``accidental'' degeneracy appears featuring an $8\,e^2/h$ first quantized step due to the mexican-hat shape of the gapped band structure in BLG. The peculiar eightfold degeneracy is also observed in the depopulation of magnetoelectric subbands (see Fig.\,\ref{fig:S9}). Unlike the data presented in the main text, here the first Landau level fully emerges from the first size-quantized energy level of the QPC. This trend can be captured within our model by shifting the size-quantized energy levels $E_N^\mathrm{QPC}\rightarrow\, E_{N-1}^\mathrm{QPC}$ for $N\geq 1$.\\

\section{Model: Landau levels in BLG QPC}

\subsection{Simplest model}

Here, we present details of the simplified model used in the main text 
to describe qualitatively the magnetoelectric subbands. 
This model disregards the effects related to the fine structure of the BLG spectrum 
(trigonal warping and mexican-hat features) \cite{Knothe2018} and the Zeeman splitting.
In this approximation, the $2\times 2$ Hamiltonian \cite{McCann2006} for the lowest conductance and valence bands in valleys $K$ and $K'$ can be written in the following form:
	\begin{gather}
		\hat{H}_K = \left( 
		\begin{matrix}
			\Delta && \frac{1}{2m}(\hat{p}_x - i \hat{p}_y)^2 \\
			\frac{1}{2m}(\hat{p}_x + i \hat{p}_y)^2 && -\Delta
		\end{matrix} 
		\right),\\
		\hat{H}_{K'} = \left( 
		\begin{matrix}
			\Delta && \frac{1}{2m}(\hat{p}_x + i \hat{p}_y)^2 \\
			\frac{1}{2m}(\hat{p}_x - i \hat{p}_y)^2 && -\Delta
		\end{matrix} 
		\right).
	\end{gather}
Here $m$ is the effective mass in BLG, $\Delta$ is half of the gap introduced by the displacement field, \textit{i.e.}, the difference between on-site energies of upper and lower graphene layers. This Hamiltonian acts in the space of sublattices A1 and B2 of the two layers.

To describe the main features of the QPC conductance in a magnetic field, we consider an infinite BLG strip of width $2W$ in the $x$--$y$ plane, with the $y$ axis oriented across the strip. 
This model corresponds to an infinite gap outside the strip, and we use below the zero boundary
conditions for the eigenfunctions at $y=\pm W$.
The magnetic field $B$ is included in the Hamiltonian via the minimal coupling 
$\hat{\bf{p}} \rightarrow \hat{\bf{p}} - e {\bf{A}}/c$, where for the vector potential we use the gauge ${\bf{A}} = (-By,\, 0,\, 0)$.
It is convenient to introduce the following notation 
\begin{gather}
	\hat{p}_+\!=\hat{p}_x-\frac{e}{c}A_x+i\hat{p}_y-i\frac{e}{c}A_y=
	-i\hbar\frac{\partial}{\partial x}-\hbar \left(\frac{y}{l_B^2}-\frac{\partial}{\partial y}\right), \notag 
	\\
	\hat{p}_-\!=\hat{p}_x-\frac{e}{c}A_x-i\hat{p}_y+i\frac{e}{c}A_y=
	-i\hbar\frac{\partial}{\partial x}-\hbar\left(\frac{y}{l_B^2}+\frac{\partial}{\partial y}\right), \notag
\end{gather}
where $l_B = \sqrt{\hbar c / e B}$ is the magnetic length and the commutator of $p_\pm$
is nonzero in a finite magnetic field:
\begin{equation}
	\left[\hat{p}_-,\, \hat{p}_+ \right] = \frac{2 \hbar^2}{l_B^2}.
\end{equation}

In the absence of magnetic field, the conductance steps are determined by the size-quantization levels at zero momentum along the strip.
In strong magnetic fields, $l_B \ll W$ , the conductance is determined by the edge states corresponding to the bulk Landau levels, and we again can set $k_x = 0$ to obtain the steps in conductance.    
The spectrum for $k_x=0$ is found from the following equations for the two valleys:
\begin{gather}
	\label{K1eq}
	K: \quad
	\begin{cases}
		\left(E^2 - \Delta^2 - \dfrac{\hat{p}_-^2 \hat{p}_+^2}{4m^2}  \right)\!\psi_{A1}(y)=0, 
		\\[0.2cm]
		\psi_{A1}(\pm W) = 0, \quad \psi_{B2}(\pm W) = 0, 
	\end{cases} \\[0.5cm] 
	\label{K2eq}
	K': \quad
	\begin{cases}
		\left(E^2 - \Delta^2 - \dfrac{\hat{p}_+^2 \hat{p}_-^2}{4m^2}  \right)\!\psi_{A1}(y)=0, 
		\\[0.2cm]
		\psi_{A1}(\pm W) = 0, \quad
		\psi_{B2}(\pm W) = 0. 
	\end{cases} 	 
\end{gather}
Introducing the operator
\begin{equation}
	\hat{A} = \frac{1}{2m} \hat{p}_- \hat{p}_+ = \frac{1}{2m} \hbar^2 \left( - \frac{\partial^2}{\partial y^2} + \frac{y^2}{l_B^4} + \frac{1}{l_B^2} \right),
\end{equation}
we re-write equations for $\psi_{A1}$ as follows:
\\
\begin{gather}
	K: \, \left[ E^2 - \Delta^2 - \left(\hat{A} + \frac{\hbar^2 }{2m l_B^2}\right)^2\!
	+\left(\frac{\hbar^2}{2m l_B^2}\right)^2 \right]\psi_{A1}\!=\!0\,, \label{K1eqSimple}\\
	K': \, \left[ E^2 - \Delta^2 - \left(\hat{A} - \frac{3\hbar^2}{2m l_B^2}\right)^2\!
	+\left(\frac{\hbar^2 }{2m l_B^2}\right)^2 \right]\psi_{A1}\!=\!0\,. \label{K2eqSimple}
\end{gather}

It is convenient to introduce the auxiliary eigenfunctions and eigenenergies of the operator 
$\hat{A}$ with zero boundary conditions, 
\begin{gather}
	\label{Aeq}
	\hat{A} \varphi_n = \varepsilon_n^A \varphi_n\,, \quad
	\varphi_n(\pm W) = 0\,.
\end{gather}
The functions $\varphi_n$ form a complete basis and every function satisfying the zero boundary conditions at $\pm W$ can be decomposed over this basis:
\begin{equation}
	\psi_{A1}^{(n)} = \sum_m a_m^{(n)} \varphi_m\,, \qquad
	\psi_{B2}^{(n)} = \sum_m b_m^{(n)} \varphi_m\,.
\end{equation}
One then substitutes $\psi_{A1}^{(n)}$ in Eqs.\ (\ref{K1eqSimple}) and (\ref{K2eqSimple}), which
yields the eigenenergies (we consider here only positive energies)
\begin{gather}
	E_n^K 
	= \sqrt{\Delta^2 + \left( \varepsilon_n^A + \hbar \omega_B/2 \right)^2 - (\hbar \omega_B/2)^2 }\,, \\
	E_n^{K'}
	= \sqrt{\Delta^2 + \left( \varepsilon_n^A -3 \hbar \omega_B/2 \right)^2 - (\hbar \omega_B/2)^2 }\,,
\end{gather}  
where $\hbar \omega_B = \hbar^2/m l_B^2$ is the cyclotron energy. Note that the shifts of
$\varepsilon_n^A$ for $K$ and $K'$ are different. This leads to the valley splitting by the magnetic field.
The coefficients in the expansion of the eigenfunctions $\psi_{A1}$ and $\psi_{B2}^{(n)}$ in $\varphi_m$
are given by
\begin{equation}
	a_m^{(n)} = \delta_{nm}, \quad b_m^{(n)}=\frac{\langle m |\hat{p}_+^2|n\rangle}{E_n^{K,K'}+\Delta},
\end{equation}
where $\delta_{nm}$ is the Kronecker delta symbol.\\

In the absence of magnetic field, $B=0$, one expresses the 
size-quantization levels of the QPC,
\begin{gather}
	\label{Ek1k2B0}
	E_n^{K,K'} = \sqrt{\Delta^2 + \left[E_n^{(0)}\right]^2},
\end{gather}
through the energy levels in a quantum well with infinitly high walls, 
$E_n^{(0)}=\varepsilon_n^A(B=0)$. 
The zero-$B$ levels $E_n^{K,K'}$ of the QPC are degenerate in the valleys. 
In strong magnetic fields, $l_B \ll W$, the boundary conditions are not important (for $k_x = 0$) and one can use the same procedure to obtain the Landau levels in BLG:
\begin{gather}
	\hat{A} \tilde{\psi}_{A1}^n = \frac{\hbar^2}{m l_B^2} (n + 1) \tilde{\psi}_{A1}^n, \\
	E_n^{K} = \sqrt{\Delta^2 + \left(\frac{\hbar^2}{m l_B^2}\right)^2 (n+1)(n+2)}, \\
	E_n^{K'} = \sqrt{\Delta^2 + \left(\frac{\hbar^2}{m l_B^2}\right)^2 (n-1)n}.
\end{gather}

To find the energy levels at intermediate magnetic fields, one needs to solve exactly the problem defined by Eqs.\ (\ref{K1eq}) and (\ref{K2eq}), which reduces to Eq. (\ref{Aeq}). However, even without finding the exact energy levels, one sees that the size-quantized wave functions with the given number $n$ transforms into the harmonic oscillator wave functions with the same number, i.e., 
$\psi_{A1}^n \xrightarrow[B \rightarrow \infty]{} \tilde{\psi}_{A1}^{n}$. 
In order to describe the experimental data, we used the following simplest interpolation 
formulas:
\begin{gather}
	\varepsilon_n^A = \sqrt{\varepsilon_n^2  + (\hbar \omega_B)^2(n+1)^2}, \\
	\varepsilon_n = \sqrt{\left(E_n^\text{QPC}\right)^2 + 2 \Delta E_n^\text{QPC}}.
\end{gather}  
Here, $\Delta$ is used as a free fitting parameter, whereas $E_n^\mathrm{QPC}$ is extracted from source-drain bias spectroscopy as described above. For $\Delta$ we have used $35\,$meV in Fig.\,4d of the main text and $18\,$meV in Fig.\ 9b of the SM. This is why the model is considered as semi-phenomenological. Indeed, in order to describe the energy levels in a fully analytical manner, one should find the actual profile of the constriction, which is determined by the electrostatic properties of the setup. 

\subsection{From four- to eight-fold degeneracy}

\begin{figure*}
		\centering
		\includegraphics[width=0.41\textwidth]{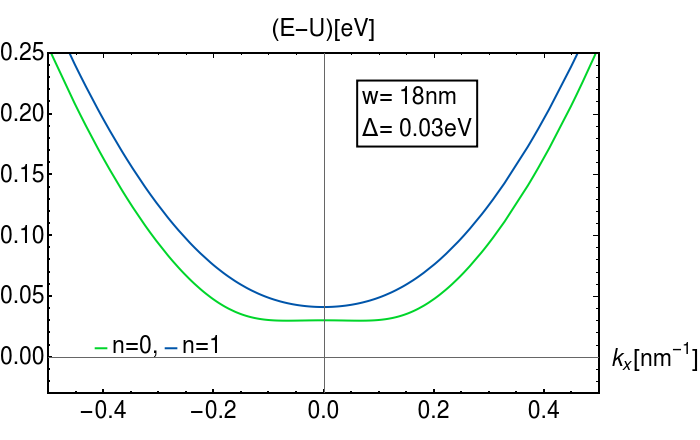}\hspace*{0.5cm}
		\includegraphics[width=0.41\textwidth]{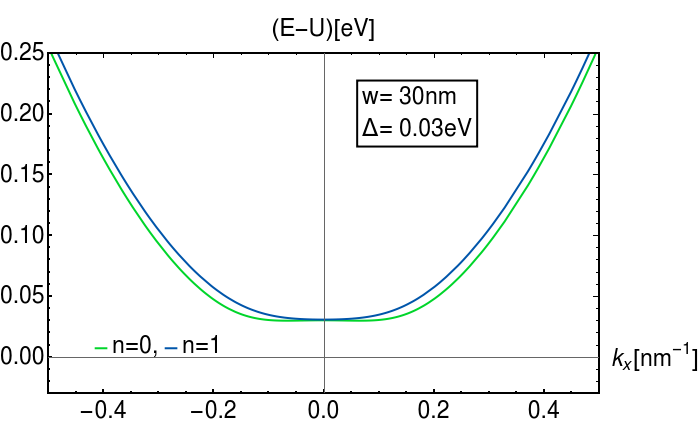}\vspace*{0.5cm}
		\includegraphics[width=0.41\textwidth]{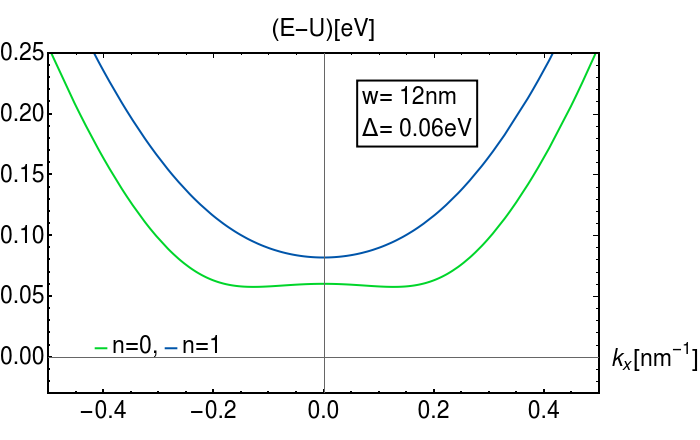}\hspace*{0.5cm}
		\includegraphics[width=0.41\textwidth]{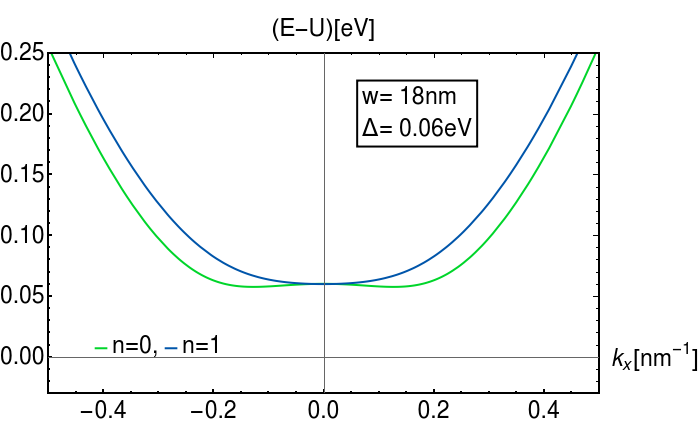}
		\caption{Evolution of zero-$B$ energy bands in a QPC with increasing width of the QPC for
for the two values of $\Delta$.}
		\label{fig:S10}
	\end{figure*}

Here, we briefly discuss the additional ``accidental'' degeneracy of the conductance related to the mexican-hat feature of the BLG spectrum \cite{Knothe2018}.
The starting point for the effective two-band Hamiltonian \cite{McCann2006} is the tight-binding Hamiltonian,
\begin{align*}
	\hat{\mathcal{H}}=\begin{pmatrix}
		\epsilon_{A1} & -\gamma_0f(\vec{k}) & \gamma_4f(\vec{k}) & -\gamma_4 f^\star(\vec{k})\\
		\gamma_0f^\star(\vec{k}) & \epsilon_{B1} & \gamma_1 & \gamma_4f(\vec{k})\\
		\gamma_4 f^\star(\vec{k}) & \gamma_1 & \epsilon_{A2} & -\gamma_0 f(\vec{k})\\
		-\gamma_3 f/\vec{k}) & \gamma_4 f^\star(\vec{k}) & -\gamma_0 f^\star(\vec{k}) & \epsilon_{B2}
	\end{pmatrix},
\end{align*}
which acts on the orbital states $A1, B1, A2, B2$.
Expanding $f(\vec{k})=\exp(ik_ya/\sqrt{3}) + 2 \exp(-ik_ya/2\sqrt{3})\cos(k_xa/2)$ around 
$\vec{K}_\pm=\pm(+4\pi/3a,0)$ and omitting both the $\gamma_4$ and the trigonal warping 
induced by $\gamma_3$, one obtains the effective four-band Hamiltonian
\begin{align}
	\hat{\mathcal{H}}_4=
	\begin{pmatrix}
		\epsilon_{A1} & v\pi^\dagger & 0 & 0\\
		v\pi & \epsilon_{B1} & \gamma_1 & 0\\
		0 & \gamma_1 & \epsilon_{A2} & v\pi^\dagger\\
		0 & 0 & v\pi & \epsilon_{B2}
	\end{pmatrix},\qquad 
	\pi:=\xi p_x+i p_y,
\end{align}
where $v=\sqrt{3} a \gamma_0/2\hbar$ and $\xi=\pm 1$ labels the two valleys.

The QPC channel is modelled by the $y$-dependence of the sublattice energies:
\begin{align}
	\epsilon_{A1}=\epsilon_{B1}=U(y)+\Delta(y),\\
	\epsilon_{A2}=\epsilon_{B2}=U(y)-\Delta(y).
\end{align}
In order to simplify the calculations, we further expand the $4\times 4$ Hamiltonian to get 
an effective two-band Hamiltonian for the low energy components $(\psi_{A1}, \psi_{B2})$. 
Assuming the step-like change of $U(y)$ and $\Delta(y)$ forming the channel,
and $U(y)=U$ and $\Delta(y)=\Delta$ for $|y|<W$,
the  $2\times 2$ Hamiltonian in the channel region is given by
\begin{gather}
	\hat{\mathcal{H}}_2
	=\begin{pmatrix}
		U+\Delta & \ -\dfrac{\pi^{\dagger 2}}{2m} \\
		-\dfrac{\pi^2}{2m}  & \ U-\Delta
	\end{pmatrix}+
	2\Delta \frac{v^2}{\gamma_1^2}\begin{pmatrix}
		\pi^\dagger\pi & 0\\
		0 & -\pi\pi^\dagger
	\end{pmatrix}.
\end{gather}
In the simplest model above, we have neglected the second term proportional to $\Delta v^2/\gamma_1^2$. 
Below, we analyze the effect of this term.

With the same boundary conditions as above, we solve the Schr\"odinger equation
\begin{gather}
	\mathcal{H}_2 \begin{pmatrix}
		\psi_{A1}\\\psi_{B2}
	\end{pmatrix} = E\begin{pmatrix}
		\psi_{A1}\\\psi_{B2}
	\end{pmatrix}, \notag\\ 
	\psi_{A1}(y= \pm W)=\psi_{B2}(y= \pm W)=0.
\end{gather}
Expressing $\psi_{B2}$ through $\psi_{A1}$, 
we get
\begin{widetext}
	\begin{align}
		\left[ (E-U)^2 - \Delta^2-\frac{1}{(2m)^2}(\pi^\dagger\pi)^2+4\Delta^2 \frac{v^2}{\gamma^2} \pi^\dagger\pi -4\Delta^2 \frac{v^4}{\gamma_1^4}(\pi^\dagger\pi)^2\right]\psi_{A1}=0.
	\end{align}
Without magnetic fields, all momentum operators commute and $\pi^\dagger\pi=p_x^2+p_y^2$. Thus this differential equation can be solved by the general ansatz
\begin{align}
	\psi_{A1}(x,y)\propto \exp(\pm ik_x x)\exp(\pm ik_y y).
\end{align}
In principle, one should consider linear combinations with all four possibilities for the signs, but since there is no restriction along the $x$ direction, it is reasonable to chose even waves with one sign along this direction. Using this ansatz we get the following condition on the momenta
	\begin{align}
		(E-U)^2 - \Delta^2-\frac{\hbar^4}{(2m)^2}(k_x^2+k_y^2)^2+4\Delta^2 \hbar^2\frac{v^2}{\gamma^2} (k_x^2+k_y^2)-4\Delta^2\hbar^4 \frac{v^4}{\gamma_1^4}(k_x^2+k_y^2)^2=0.
	\end{align}
To account for the boundary conditions along the $y$ direction,
we introduce quantized momenta and the corresponding discrete energies
\begin{align}
	k_y = \frac{n \pi}{2 W},\quad E_n^{(0)} =\frac{\hbar^2}{2m} \frac{n^2\pi^2}{4W^2},\quad  n\in \mathbb{Z},
\end{align}
yielding the continuous energy spectrum $E_n(k_x)$ in the form
	\begin{align}
		(E_n-U)^2 = \Delta^2+4\Delta^2\frac{v^4}{\gamma_1^4}(2mE_n^{(0)} +\hbar^2 k_x^2)^2 - 4\Delta^2 \frac{v^2}{\gamma_1^2}\left(2mE_n^{(0)}+\hbar^2k_x^2\right)+\frac{1}{(2m)^2}\left(2mE_n^{(0)} +\hbar^2k_x^2\right)^2
	\end{align}

In sufficiently wide channels, when the two conditions 
\begin{align}
	W>\frac{\pi \hbar v}{2 \gamma_1}\quad \text{and} 
	\quad \Delta >\frac{\pi \hbar \gamma_1^2}{4 m v} \frac{1}{ \sqrt{4 W^2 \gamma _1^2-\pi ^2 \hbar^2 v^2}},
\end{align}
are simultaneously satisfied, the two lowest bands ($n=0$ and $n=1$) touch at 
\begin{align}
	k_x=\pm\frac{\sqrt{16\Delta ^2 m^2 v^2 \left(4 \gamma _1^2 W^2 -\pi ^2\hbar^2 v^2\right)-\pi ^2\hbar^2 \gamma _1^4}}{2\sqrt{2} W \hbar \sqrt{16 \Delta ^2 m^2 v^4+\gamma _1^4}}.
\end{align}

\end{widetext}

It is this touching of the bands \cite{Knothe2018}, see Fig. \ref{fig:S10}, that leads to the additional degeneracy
of the lowest conductance steps, replacing the generic four-fold degeneracy by the eight-fold degeneracy in wide constrictions.